\documentclass[12pt]{article}
\usepackage{amsmath}
\usepackage{latexsym}
\usepackage{bbm}

\def\a{\alpha}

\def\th{\theta}

\def\l{\lambda}
\def\m{\mu}
\def\n{\nu}

\def\r{\rho}

\def\s{\sigma}
\def\t{\tau}

\def\pa{\partial}

\def\and{{\rm and}}

\def\ie{{\it i.e.,} }

\def\IC{\mathbbm C}

\def\IR{\mathbbm R}
\def\IZ{\mathbbm Z}

\begin{document}
\vspace*{-1.0in}
\thispagestyle{empty}
\begin{flushright}
CALT-68-2947
\end{flushright}
\baselineskip = 18pt
\parskip = 6pt

\vspace{1.0in} {\Large
\begin{center}
Multicharge Superstrings\footnote{This paper is an expanded version of the opening lecture given at Strings 2013 in Seoul.}\end{center}} \vspace{.25in}

\begin{center}
John H. Schwarz
\\
\emph{California Institute of Technology\\ Pasadena, CA  91125, USA}
\end{center}
\vspace{.25in}

\begin{center}
\textbf{Abstract}
\end{center}
\begin{quotation}
\noindent

There are an infinite number of type IIB superstrings in ten dimensions, called $(p,q)$ strings, that are labeled by two distinct string charges. They can form string junctions and string networks. These are the key to understanding the Coulomb branch BPS spectrum of ${\cal N} =4$ super Yang--Mills theories in four dimensions. Nongravitational theories that are analogous to type IIB superstring theory in some respects are the ADE $(2,0)$ six-dimensional theories on the Coulomb branch. Each of them contains a finite number of half-BPS superstrings, which can be labeled by the root vectors of the Lie algebra. When the rank is greater than one, these strings can also form string junctions and string networks. The relationship between ${\cal N} =4$ super Yang--Mills theories in four dimensions and the $(2,0)$ theories in six dimensions can be explored by utilizing the duality between type IIB superstring theory compactified on a circle and M-theory compactified on a torus.

\end{quotation}

\newpage

\pagenumbering{arabic}

\tableofcontents

\newpage

\section{Introduction}

In the mid 1990s it was recognized that the $SL(2, \IR)$ symmetry of type IIB supergravity is broken to an $SL(2,\IZ)$ duality symmetry in type IIB superstring theory \cite{Hull:1994ys}. Moreover, the latter has two distinct string charges leading to an infinite family of $(p,q)$ superstrings that form an irreducible multiplet of the $SL(2,\IZ)$ duality symmetry \cite{Schwarz:1995dk}, and there exist three-string junctions that can be used to build string networks. The first part of this paper reviews the basic facts about these strings as well as the nonperturbative BPS spectrum obtained after compactification of type IIB superstring theory on a circle. The dual description in terms of M-theory compactified on a torus is also discussed. Most of this was understood in the 1990s, but some issues were only recognized more recently.

The second part of this paper examines the six-dimensional $(2,0)$ theories associated with a
set of parallel flat M5-branes. These theories, which have an ADE classification, can also be understood as arising in type IIB superstring theory at an ADE singularity \cite{Witten:1995zh}. Such a theory on the Coulomb branch has a finite set of self-dual superstrings. For an ADE Lie algebra of rank $r$ the number of distinct string charges is $r$ and the distinct BPS strings correspond to the roots of the Lie algebra. The six-dimensional self-dual strings also form three-string junctions that can be used to build string networks.


These facts do not constitute a complete formulation of the $(2,0)$ theories, even on the Coulomb branch. Rather, they provide an effective way of characterizing the various BPS brane configurations in the theory, both before and after toroidal compactification, analogous to what one can do in type IIB superstring theory. The understanding of the $(2,0)$ theories is more limited than the type IIB theory in that the six-dimensional theory contains no analog of the dilaton that can be tuned to make one of the ten-dimensional superstrings perturbative. In this respect it is more like M-theory. Also, it is unclear how to formulate the $(2,0)$ theories in the conformal limit in which the M5-branes become coincident.

\section{M/IIB duality}

One of the dualities discovered in the 1990s relates M-theory compactified on a torus to type IIB superstring theory compactified on a circle \cite{Aspinwall:1995fw,Schwarz:1995jq}. I will now describe how the parameters associated with each of these are related. Let us first denote the Planck length for M-theory in eleven dimensions by $l_M$ and the Planck length for type IIB superstring theory in ten dimensions by $l_B$. The latter is related to the string scale, denoted $l_S$, and the string coupling constant, denoted $g_S$, by
\begin{equation}\label{1a}l_B = g_S^{1/4} l_S.\end{equation}
One way of understanding the exponent of $g_S$ is to recall that Newton's constant in ten dimensions is proportional to $l_B^8$ and that it is proportional to $g_S^2$ when reexpressed in terms of $l_S$.

Let us characterize the torus on which M-theory is compactified by a modular parameter, $\t_M$,
and an area, $(2\pi)^2 A_M$. This normalization will minimize the appearance of factors of $2\pi$
in subsequent equations. Correspondingly, the type IIB theory has a discrete $SL(2,\IZ)$ duality
symmetry under which the complex scalar field $\l = C_0 +i e^{-\Phi}$ transforms in the usual
nonlinear manner, $\l \to (a\l +b)/(c\l +d)$. The real scalar fields $C_0$ and $\Phi$ are the RR scalar and the dilaton. Let us denote the expectation value of $\l$ by
\begin{equation} \label{1b} \t_B  = \langle \l \rangle = \frac {\theta}{2\pi} + \frac {i} {g_S}. \end{equation}
In this paper we only consider backgrounds in which no 7-branes are present, and $\t_B$ is position independent.

Identifying the $SL(2,\IZ)$ duality group of the type IIB theory with the modular group of the M-theory torus, we obtain
\begin{equation}\label{1c} \t_M = \t_B. \end{equation}
Therefore in the following we drop the subscripts and simply write $\t= \t_1 + i \t_2$ in both cases. It is an important fact that the Einstein-frame metric of type IIB
superstring theory is $SL(2,\IZ)$ invariant. It follows that the Planck length $l_B$ is also
$SL(2,\IZ)$ invariant, but the string scale $l_S$ is not invariant.

\subsection{Matching 2-branes and 3-branes}

Further relations are obtained by identifying various supersymmetric brane configurations and equating tensions \cite{Schwarz:1995jq,Schwarz:1996bh}. To do this, first we need to recall that the tensions of the supersymmetric M-branes are
\begin{equation}\label{1d} T_{M2} = \frac {2\pi}{(2\pi l_M)^3} \quad {\rm and} \quad T_{M5} = \frac {2\pi}{(2\pi l_M)^6}. \end{equation}
Also, the tension of the type IIB D3-brane is given by the $SL(2,\IZ)$ invariant formula
\begin{equation}\label{1e} T_{D3} = \frac {2\pi}{(2\pi l_S)^4 g_S} = \frac {2\pi}{(2\pi l_B)^4}. \end{equation}
Equating the tension of an M5-brane wrapped on the torus with the tension of a D3-brane gives
\begin{equation}\label{1f} l_B^4 = l_M^6/A_M. \end{equation}
Similarly, equating the tension of a D3-brane wrapped on a circle of radius $R_B$ with the
tension of an M2-brane gives
\begin{equation}\label{1g} l_M^3 = l_B^4/R_B. \end{equation}
Equations (\ref{1f}) and ({\ref{1g}) can be combined to give the following relation between dimensionless quantities:
\begin{equation}\label{1h} (A_M/l_M^2)^3 = (l_B/R_B)^4. \end{equation}
In this sense one could say that $A_M \sim R_B^{-4/3}$. On the other hand, if we were to set $l_M =1$, we would conclude that $A_M = R_B^{-1}$. Alternatively, if we were to set $l_B =1$, this would lead to $A_M = R_B^{-2}$. Therefore it is less confusing to keep both of these scales arbitrary. This completes the matching of 2-branes and 3-branes in nine dimensions.

\subsection{Matching 1-branes}

It is also instructive to relate the various BPS 0-branes and 1-branes that exist in nine dimensions even though their matching give no further relations among the parameters. The type IIB
theory has an infinite spectrum of half-BPS $(p,q)$ strings in ten dimensions, which can be regarded as bound states of $p$ fundamental strings and $q$ D-strings. $p$ and $q$ are integer charges that measure the strengths with which the strings couple to the two-form gauge fields $B_2$ and $C_2$.
These strings are stable whenever $p$ and $q$ are coprime.  Their tensions are given by \cite{Schwarz:1995jq}
\begin{equation}\label{1i} T_{p,q} = |p + q\t|T_F, \end{equation}
where $T_F$, the tension of the fundamental string, is
\begin{equation}\label{1j} T_F = \frac{1}{2\pi l_S^2} = \frac{1}{2\pi \sqrt{\t_2} l_B^2}. \end{equation}
The $\t$ dependence of these formulas is consistent with the fact that the string charges $p$ and $q$ transform linearly under $SL(2,\IZ)$ transformations. This is required because the 2-form gauge fields also transform as a doublet.

After compactification on a circle, the $(p,q)$ strings continue to exist in the nine noncompact dimensions. Their tensions are reproduced by wrapping an M2-brane on a geodesic curve in the $(p,q)$ homology class of the M-theory torus. The length of such a cycle is $2\pi |p+q\t|\sqrt{A_M /\t_2}$.
Multiplying by $T_{M2}$ and using the identities above gives perfect agreement with $T_{p,q}$.

As an aside, note that the tension of the fundamental string is proportional
to $\sqrt{1/\t_2} = \sqrt{g_S}$ when measured in Planck units, \ie in the ``Einstein frame.'' These
are the natural units for exhibiting the $SL(2,\IZ)$ symmetry. This means that in perturbation
theory, when one expands about $g_S =0$, one is describing tensionless strings from the point of view of the Einstein frame.  The way to avoid thinking about tensionless strings, which are very problematic, is to utilize the string frame, where the fundamental string tension remains finite even as $g_S \to 0$. This has the added benefit of decoupling all nonperturbative excitations, which shouldn't be included in perturbative calculations. Of course, string theorists have been doing this from the beginning -- long before they appreciated this issue.

\subsection{Matching 0-branes}

The analysis of BPS 0-branes in nine dimensions is a bit more involved, because they can carry three types of charges. Let us start with the simplest case. An M2-brane wrapped $K$ times on the torus is dual to $K$ units of Kaluza--Klein (KK) excitation of the type IIB circle. This gives half-BPS states with mass
\begin{equation}\label{1k} M_K = |K|/R_B = |K| A_M/l_M^3. \end{equation}
The agreement of the two expressions is a consequence of formulas given previously.

It is interesting to consider the supermultiplet structure of these excitations. The little group that classifies massless excitations in $D$ dimensions, $Spin(D-2)$, is the same as the one that classifies massive excitations in $D-1$ dimensions. In fact, each of the KK excitations on a circle of a $D$-dimensional massless supermultiplet is interpreted as a massive supermultiplet in $D-1$ dimensions that is identical to the massless one in $D$ dimensions. Thus, the KK excitations of the
ten-dimensional type IIB supergravity multiplet correspond to massive chiral supermultiplets in nine dimensions. In particular, they contain two massive gravitinos of the same chirality. This may seem surprising from the M-theory perspective, since eleven-dimensional supergravity does not contain any chiral fields. Nonetheless, the duality implies that this chirality is a necessary feature of the BPS spectrum of nine-dimensional 0-branes that originate as wrapped M2-branes.

A second class of half-BPS states arises from a $(p,q)$ string wrapped $W$ times on the type IIB circle, which is dual to an $(n_1, n_2)$ KK excitation on the M-theory torus. The masses are
\begin{equation} \label{1l}
M_{n_1,n_2}  = 2\pi R_B |W| T_{p,q}= \frac{|n_1 + n_2 \t|} {\sqrt{\t_2 A_M}}. \end{equation}
The two expressions agree using identities obtained previously for the identification
\begin{equation}  \label{1m}(n_1, n_2) = (Wp, Wq). \end{equation}
Since $p$ and $q$ are coprime, this means that $W$ is the greatest common divisor of $n_1$ and $n_2$.\footnote{A $(p,q)$ string with winding number $W$ is equivalent to a $(-p, -q)$ string with winding number $-W$.} In this case the massive supermultiplets in nine dimensions are identical to the massless ten-dimensional type IIA supergravity multiplet, which
is the dimensional reduction of the 11-dimensional supergravity multiplet. Thus, we conclude that
the nine-dimensional massive supermultiplets that arise as BPS modes of type IIB superstrings wrapped on a circle must be identical to the massless ten-dimensional type IIA supergravity multiplet. In particular, they are nonchiral.

It is clear from the preceding discussion that there are two distinct classes of half-BPS 0-brane
massive supermultiplets in nine dimensions. They are distinguished by which central charges of the
nine-dimensional supersymmetry algebra determine the mass: $n_1$ and $n_2$ in one case and
$K$ in the other case. A supersymmetric 0-brane in nine dimensions that carries both types of charges,
with $WK\neq 0$, can only be quarter-BPS.
This is most easily understood from the type IIB viewpoint, where we know that a fundamental string
with winding number $W$ and KK excitation number $K$ satisfies the level-matching condition
$N_L - N_R = WK$. Here $N_L$ and $N_R$ denote excitation numbers of left-moving and right-moving oscillators in the usual manner. Because of the $SL(2,\IZ)$ symmetry, this formula applies to all of the $(p,q)$ strings. Half-BPS states are given by $N_L = N_R=0$, which requires that either $W=0$ (and hence $n_1 = n_2 =0$) or $K=0$. Quarter-BPS states are given by $N_L=0$ or $N_R=0$, but not both. The appropriate choice is determined by the sign of $WK$. Altogether,
this means that the quarter-BPS states are given by exciting the left or right set of oscillators
to level $N=|WK|$. The masses of these particles are then given by
\begin{equation} \label{1n} M_{K,n_1, n_2} = M_K + M_{n_1, n_2}, \end{equation}
where $M_K$ and $ M_{n_1, n_2}$ are given in eqs. (\ref{1k}) and (\ref{1l}).

\subsection{Nine-dimensional black holes?}

We conclude this section by exploring whether there is a black-hole interpretation of the quarter-BPS 0-branes when the charges are large. For large $N=|WK|$ the number of quarter BPS states is roughly $d_N \sim \exp(2\pi\sqrt{N})$. If there is a black-hole interpretation, the entropy should be $S= 2\pi \sqrt{WK}$. The problem is that the integer $W$, the greatest common divisor of $n_1$ and $n_2$, jumps around quite wildly as the charges $n_1$ and $n_2$ are varied even though the mass does not. This is
a disturbing conclusion if one expects to find a macroscopic geometric description of these quarter-BPS ``black holes''. The basic mathematical issue is that the charges $(n_1, n_2)$ transform as a doublet of $SL(2,\IZ)$, and the greatest common divisor $W$ is the {\em only} invariant that can be constructed out of such a doublet. Thus any improvement of the entropy formula, such as including the contributions of multicentered black holes, must still give a function of $W$, and therefore it would still have the same behavior.

In the familiar examples of BPS black holes with finite area horizons in 4d and 5d, for which the duality groups are $E_7(\IZ)$ and $E_6(\IZ)$,
this problem does not arise. The entropy formulas at leading order are functions of quartic and cubic invariants that are continuous in the charges. The reason this is possible in these cases is that the invariants that appear in the entropy formulas respect the continuous $E_{7,7}$ and $E_{6,6}$ symmetries -- not just their discrete subgroups. There are subleading corrections to the entropy in these cases that only have the symmetry of the discrete subgroup, and they do jump around in the manner that we are finding in nine dimensions. However, they give very tiny corrections to the entropy, which do not require a macroscopic description.

One possibility for avoiding this conclusion would be to suppose that these nine-dimensional quarter-BPS states do not exist. This is conceivable because they are at threshold for decay into a pair of half-BPS states, and the existence or nonexistence of threshold bound states is always a subtle matter. However, the fact of the matter is that these states do exist, and so we must conclude that they have no macroscopic geometric description in terms of extremal supersymmetric black holes.\footnote{I am grateful to C. Vafa for a discussion of this issue.}

\subsection{String junctions and string networks}

The $(p,q)$ strings can form three-string junctions \cite{Schwarz:1996bh}. The rules are quite simple: The string charges should be conserved at the junction, just like momenta at a vertex in a Feynman diagram. Furthermore, the angles between the strings are determined by requiring the forces given by the
string tensions to balance at the junction. One can then use these junctions to build up string networks. Again, the pictures are reminiscent of Feynman diagrams, though the meaning is different. One of these diagrams describes a string network embedded in the spatial dimensions at a given instant of time. All of these considerations concern static configurations. The three-string junction is quarter-BPS. The same is true for any string network that is two-dimensional. Networks that span more dimensions have less supersymmetry. By studying such networks, with the external strings ending on D3-branes, one can carry out detailed explorations of the spectrum of MSYM states\footnote{MSYM stands for maximally supersymmetric Yang--Mills, which is ${\cal N} =4$ in the case of four dimensions.} which preserve various amounts of supersymmetry (half-BPS, quarter-BPS, etc.) \cite{Bergman:1997yw}.

From these considerations, it is apparent that only half-BPS states occur in the $SU(2)$ theory, since there are just two D3-branes, but states with less supersymmetry occur in all of the higher-rank theories. More specifically, the positions of three D3-branes in the transverse six dimensions define a two-dimensional plane (unless they are collinear), and then quarter-BPS webs that connect them are possible. For a recent detailed discussion of quarter-BPS states in four-dimensional MSYM theories, based on string networks, see \cite{Sen:2012hv}. Among other things, this analysis provides an intuitive geometric interpretation of wall-crossing phenomena.

\section{Branes within branes}

Sometimes we are interested in considering BPS branes that contain lower-dimensional branes
embedded within them in such a way that the charge of the lower-dimensional brane is uniformly
distributed within the higher-dimensional brane. As we will explain, in many cases the smearing of the charge can be interpreted in terms of localized objects from a higher-dimensional viewpoint. When this is the case, the energy density of the brane is no longer equal to its tension.\footnote{I am grateful to J. Maldacena for reminding me that tension and energy density are not the same thing.}

\subsection{Strings}

Let us begin with the example of the nine-dimensional $(p,q)$ strings. A more general possibility, which is still half BPS, is for a $(p,q)$ string to wrap the circular dimension in a helical manner. In other words, the coordinate of the string in a noncompact spatial direction is proportional to the angular coordinate of the string on the circle so that the string has a constant winding number per unit length $\r$. The string remains localized in the seven transverse spatial dimensions. Such a curve is a straight line in the covering space, and therefore it is still half BPS.

From the nine-dimensional viewpoint this produces a $(p,q)$
string that also has finite charge densities $(\r_1, \r_2) = \r(p,q)$. These charge densities measure the coupling of the string to the nine-dimensional one-form gauge fields $B_{\m 9}$ and $C_{\m 9}$, whereas the string charges $(p,q)$ measure the couplings to the nine-dimensional two-forms $B_{\m\n}$ and $C_{\m\n}$. When $\r =0$, this construction reduces to the nine-dimensional $(p,q)$ string discussed previously. By computing the ten-dimensional length of the string it is easy to deduce that the apparent energy density ${\cal E}$ of the string in nine dimensions is
\begin{equation} \label{2a} {\cal E}_{p,q}(\r) = \sqrt{1 + (2\pi R\r)^2} T_{p,q} = \sqrt{T_{p,q}^2 + (\r M_{p,q})^2}, \end{equation}
where $M_{p,q}= 2\pi R T_{p,q}$, as before. Alternatively, we can define
\begin{equation} \cos \phi = 1/\sqrt{1 + (2\pi R\r)^2}, \end{equation}
where $\phi$ is the angle that the string makes between the circle direction and the noncompact direction. Then the nine-dimensional energy density takes the form
\begin{equation} {\cal E}_{p,q}(\r) = T_{p,q}/\cos \phi. \end{equation}
On the other hand,  the apparent tension $ T_{p,q}(\r)$ of the string in nine dimensions is reduced by the same factor
\begin{equation}  T_{p,q}(\r) = T_{p,q}\cos \phi, \end{equation}
and the component of the tension along the circle is $T_{p,q}\sin \phi$.

Another example of a string carrying a 0-brane charge can be understood as a limiting case
of the above construction. Specifically, the type IIA superstring can be decorated with
a D0-brane charge density $\r$ per unit length. In this case the energy density is
\begin{equation} {\cal E}(\r) = \sqrt{T_{F1}^2 + (\r M_{D0})^2}, \end{equation}
where $T_{F1} = (2\pi l_s^2)^{-1}$ is the fundamental string tension and $M_{D0} =(g_s l_s)^{-1}$ is the D0-brane mass. The eleven-dimensional interpretation is clear:
the string is an M2-brane wrapped on the M-theory circle and the D0-brane charge arises
as KK excitations of the circle. However, there is a significant difference from the previous
example. Namely, the higher-dimensional perspective does not describe such a charged string
as a fundamental brane whose tension and energy density are equal.

\subsection{2-branes}

For an example of a 2-brane with smeared string charges, let us turn to the M-theory picture. Suppose one spatial coordinate of a noncompact M2-brane wraps the $(p,q)$ cycle of the M-theory torus helically, with a winding number per unit length $\r_{p,q}$, just as the string wrapped the circle in the previous example. This produces a 2-brane in nine dimensions with a constant density $\r_{p,q}$ of $(p,q)$ string charge. The nine-dimensional energy density of the 2-brane in this case is
\begin{equation} \label{2b} {\cal E}_{M2}(\r_{p,q}) = \sqrt{T_{M2}^2 + (\r_{p,q} T_{p,q})^2} = T_{M2}/\cos\phi. \end{equation}
As before, the nine-dimensional tension is $T_{M2}(\r_{p,q}) = T_{M2} \cos\phi$.

\subsection{Dyonic 3-branes}

Let us now carry out an analogous construction for the D3-brane. Specifically, one of its dimensions is wrapped on the type IIB circle with winding number $\r_{M2}$ per unit length. This results in a 3-brane with a density $\r_{M2}$ of M2-brane charge. (Recall that a wrapped D3-brane is dual to an unwrapped M2-brane.) The resulting nine-dimensional energy density is
\begin{equation} \label{2c} {\cal E}_{D3}(\r_{M2}) = \sqrt{T_{D3}^2 + (\r_{M2} T_{M2})^2} = T_{D3} /\cos \phi. \end{equation}
and the nine-dimensional tension is
\begin{equation} T_{D3}(\r_{M2}) = T_{D3} \cos \phi. \end{equation}
An interesting feature of this case is that in nine dimensions the D3-brane charge and the M2-brane charge are electric/magnetic duals. Thus, this 3-brane carries mutually non-local charges. This is no more exotic than the ten-dimensional D3-brane, which has a self-dual charge. Indeed, from a ten-dimensional viewpoint this is what it is.

Now suppose that we wish to study the BPS spectrum of the gauge theory associated to $N$ parallel D3-branes in the case that one of the six dimensions transverse to the branes is a (small) circle. From the nine-dimensional perspective, we have the following
\begin{itemize}
\item Each of the D3-branes is characterized by a position $\vec y$ in the transverse $\IR^5$ and an M2-brane charge density $\r_{M2}$.
\item Each of the strings that can end on a D3-brane is characterized by string charges $(p,q)$ and a winding number density $\r$.
\item A three-string junction involving strings labeled $(p_i, q_i, \r_i)$, $i = 1,2,3$, must satisfy string charge conservation, $ \sum p_i = \sum q_i =0$, as before. Also, the force (determined by the tensions) must balance at the junction in the five noncompact dimensions and in the compact dimension. The latter condition effectively relates the $\r$ parameters.
\end{itemize}
These considerations raise the following question: What does it mean from the MSYM theory point of view, as well as the dual M5-brane point of view, for the D3-branes to have constant M2-brane charge densities? We will return to this question later.

The system we have just described is the type IIB dual of a system of parallel M5-branes wrapped
on a torus. The duality described earlier implies that the limit in which the type IIB circle
is shrunk to zero size, corresponds to decompactification of the torus. The $\t$ parameter should be held constant in taking this limit, though the $\t$ dependence should drop out at the end. Multiple M5-brane systems are explored from other viewpoints in the next section.

\section{Multiple M5-branes}

It is an important challenge to develop a useful formulation of the $(2,0)$ six-dimensional
superconformal field theories associated with multiple coincident M5-branes. These
theories, which have an ADE classification, are in some sense the most fundamental
of all nongravitational theories. Many interesting theories in lower dimensions, as well as relations among them, have been characterized in recent years in terms of compactifications of these theories, and this has been very fruitful. This program would have a more solid foundation if we had a more explicit description of the starting point. This problem is closely connected to the even more fundamental problem of constructing a complete nonperturbative formulation of M-theory/superstring theory. The relation between the two problems is illustrated by the duality that relates M-theory on $AdS_7 \times S^4$ with $N$ units of four-form flux to the $A_{N-1}$ SCFT \cite{Maldacena:1997re}. This is a great correspondence, but both sides remain rather mysterious.

Many papers have been written over the years developing a wide range of approaches to understanding the $(2,0)$ SCFTs. Some of them certainly address important aspects of the problem. In that spirit, I wish to describe and elaborate upon a proposal that I made in 1996 \cite{Schwarz:1996pi}, which focuses on identifying the BPS excitations that are present on the Coulomb branch.

A basic fact about M5-branes is that M2-branes can end on them, much like fundamental strings
can end on D-branes. When a cylindrical M2-brane is suspended between a pair of parallel nearby
M5-branes, it can be approximated by a string within the six-dimensional world-volume
theory of the M5-branes. The tension of the string is given by the tension of the M2-brane
times the separation of the M5-branes.
This is a higher-dimension analog of a fundamental string suspended between a pair of parallel D3-branes. This analogy suggests that an understanding of the world-volume theory of a set of parallel flat M5-branes should take these strings into account. Note that the strings that enter the D3-brane analysis are transverse to the branes, whereas the ones that enter in the M5-brane analysis are along the branes.

\subsection{The (2,0) tensor multiplet}

The world-volume theory of a single M5-brane has two six-dimensional supersymmetries of the same chirality. This is analogous to the type IIB superstring theory in ten dimensions except that the M5-brane world-volume theory is nongravitational. In addition to super-Poincar\'e symmetry, the M5-brane theory has a $Spin(5) =USp(4)$ R-symmetry, which corresponds to rotations of the five transverse dimensions. The free theory has superconformal symmetry $OSp(6,2|4)$ that incorporates and extends these symmetries.

Just as in the case of a single D3-brane, there is an interacting Born-Infeld-like extension of the free theory that is not conformal \cite{Aganagic:1997zq}. That is not our interest here. In the setting when gravity and back reaction are
taken into account the distinction concerns whether the asymptotic geometry far from the branes is
flat or Anti de Sitter. In the latter case the dual field theory should be conformal, at least at
high energies. Furthermore, we wish to consider theories in which gravity is decoupled.

The massless multiplet associated with a single M5-brane consists of a 2-form field $B$ with a self-dual field strength, four symplectic-Majorana chiral spinors, and 5 scalars. In terms of the little group for massless particles, which in six dimensions is $SU(2) \times SU(2)$, the representations are
\begin{equation} (3,1) + 4(2,1) + 5(1,1). \end{equation}
Altogether, the supermultiplet contains eight bosonic modes and eight fermionic modes. The coefficients $1,4,5$ are the dimensions of irreducible representations of $USp(4)_R$. The fermion representation is hermitian (consistent with unitarity), because the ${\bf 4}$ of $USp(4)_R$ and the ${\bf 2}$ of $SU(2)$ are both pseudoreal. This is what the term ``symplectic Majorana'' refers to.

\subsection{Multiple 3-branes and $(2,0)$ theories}

As a first step towards understanding a system of multiple parallel M5-branes, let us relate
it to a dual type IIB configuration, since there are more tools available to study the dual system. With this motivation, let us once again consider the correspondence between an M5-brane wrapping the M-theory torus and an unwrapped D3-brane in the IIB theory compactified on a circle. The M5-brane is localized in the five noncompact spatial dimensions transverse to the brane, so its position is given by a five vector $\vec y$ in $\IR^5$. The dual D3-brane also has position $\vec y$ in the $\IR^5$ that is transverse to its world volume. However, it also has an angular position $\th$ on the type IIB circle. If there are $N$ parallel M5-branes, they and the dual D3-branes have five-vector positions $\vec y_i$, $i = 1,2,\ldots, N$. In addition, the D3-branes have angular positions $\th_i$. These angular positions are encoded in the M-theory picture in ``Wilson surfaces.'' These are integrals $\int_{T^2} B_i$, where the fields $B_i$, which belong to supermultiplets associated to the corresponding M5-branes, take constant values. Since these integrals only appear as phases in the path integral defining the quantum theory, the Wilson surfaces may be regarded as angular coordinates. This result is a straightforward strong-coupling generalization of the relationship between D4-branes and D3-branes under T-duality.

We can now analyze the multiple M5-brane system on a torus in terms of this dual D3-brane system.
Then the decompactification limit of the torus can be studied by shrinking the type IIB circle,  $R_B\to 0$, while holding the coordinates $\th_i$ and $\vec y_i$, as well as the modulus $\t$, fixed. In the limit the theory should be independent of $\t$ and the Wilson surfaces.
Suppose that our goal is to determine the complete BPS spectrum of stable excitations.
In the case when the $y_i$ are all distinct there is no problem determining what one ends up with in six dimensions. The BPS particle spectrum consists entirely
of $N$ massless tensor multiplets -- one for each of the M5-branes. As we saw previously, the KK
spectrum that accounts for the torus is represented in the dual picture by the winding modes of all the $(p,q)$ strings.

In the previous section we showed that a D3-brane that wraps a circular dimension helically appears
from the nine-dimensional viewpoint to have a constant M2-brane charge density embedded in it. We now see that in terms of the dual M5-brane, this means that the Wilson surface is not a constant. For example, if $x_1$ and $x_2$ refer to the torus directions and $x_3$ is a noncompact coordinate of the M5-brane, we could take $B_{12} = \a x_3$, which gives a Wilson surface linear in $x_3$ and a constant field strength $H_{123} =\a$.\footnote{Self duality of $H$ requires that $H_{045} =\a$, as well.} From the four-dimensional viewpoint $B_{12}$ is one of the six scalar fields, which we could call $\phi_6$, since the tensor supermultiplet already contains the other five. This one is a singlet of the $USp(4)_R$ symmetry. Thus, we conclude that the constant M2-brane charge density on a D3-brane is represented by a background value for $\phi_6 $ that depends linearly on the three spatial coordinates. These linear functions should be the same (except for the constant terms) for each of the $N$ D3-branes in order to keep the branes parallel and to avoid supersymmetry breaking. Then the $N-1$ differences would all be constants. The background field would only be associated with the $U(1)$ multiplet, and its rotational symmetry would be broken.

\subsection{Six-dimensional superstrings}

In order to describe a simple scenario in terms of six-dimensional superstrings that seems to account for the entire BPS spectrum, let us consider the $A_{N-1}$ theory on the Coulomb branch. It corresponds to $N$ parallel M5-branes, whose five-dimensional positions $\vec y_i$ are distinct but otherwise arbitrary. There are $N(N-1)/2$ oriented half-BPS strings associated to each pair of M5-branes $(ij)$ with $i<j$. (Interchanging $i$ and $j$ reverse the orientation.)


The zero modes of each string should form a $(2,0)$ tensor supermultiplet. In terms of physical degrees of freedom, this supermultiplet can be viewed as arising from tensoring left-moving and right-moving zero modes as follows \cite{Schwarz:1996pi}:
\begin{equation} [(2,1) +2(1,1)] \times [(2,1) +2(1,1)] =  (3,1) + 4(2,1) + 5(1,1). \end{equation}
This is analogous to the way that one forms the $(2,0)$ supergravity multiplet out of superstrings in ten dimensions. However, there are a few important respects in which the (2,0) superstring theories differ from the ten-dimensional type IIB superstring theory:
\begin{itemize}
 \item The six-dimensional theories have $N(N-1)/2$ distinct oriented strings, or twice this number if one counts the orientation reversed strings as distinct, whereas the IIB theory has an infinite number of distinct half-BPS strings (the $(p,q)$ strings). The strings in six dimensions correspond to the positive roots of the associated $A_{N-1}$ Lie algebra. The orientation reversed strings then correspond to the negative roots. This rule generalizes to $D$ and $E$ algebras. Thus, for example, the $E_8$ theory has 120 distinct strings.
 \item Ten-dimensional $(p,q)$ strings carry two distinct string charges that characterize how they couple to the two ten-dimensional two-forms $B$ and $C$. By contrast, the six-dimensional superstrings are characterized by $N-1$ charges, which characterizes their coupling to $N-1$ six-dimensional two-forms $B$. Each of the $B$ fields has a self-dual field strength, which implies that these strings have self-dual charges, like the D3-brane in ten dimensions. More generally, the number of independent charges is the rank of the Lie algebra.
 \item The ten-dimensional fundamental string has a weak coupling limit, which makes it possible to study the theory perturbatively. The six-dimensional $(2,0)$ theories do not have a weak coupling limit.
 \end{itemize}

There are $N$ 2-forms fields $B_i$ associated to the $N$ M5-branes. However, only differences are relevant and an overall `center-of mass' tensor multiplet decouples from the theory. (This is analogous to the $U(1)$ decoupling from the D3-brane theories.) Specifically, the tensor multiplet
that is the zero mode of the $(ij)$ string contains the difference $B_i - B_j$. This is also the field
for which the $(ij)$ string is a source. The charges of these strings, and hence the strings themselves, can be labeled by root vectors of the Lie algebra. The tension of the $(ij)$ string is proportional
to the separation $|\vec y_i - \vec y_j|$, because this is the distance that the M2-brane is stretched. Thus, charge conservation implies stability of all the strings unless the positions of three M5-branes are collinear. In that case there is marginal stability.

Infinitely extended straight strings, which are stable BPS objects in six dimensions, are sensible objects to consider. For such strings the zero modes, described previously, are the essential information. They carry infinite energy, of course, as well as a conserved string charge.
Since these strings are strongly coupled, there is not much to be gained by studying the excited modes of a free string --- in other words, determining the world volume conformal field theory of the string. However, when these strings wrap a compact dimension, as discussed in Sect. 4.5, the entire world volume conformal field theory becomes relevant.

\subsection{Coupling and decoupling supergravity}

Let us consider coupling $N$ massless $(2,0)$ tensor supermultiplets to six-dimensional $(2,0)$ supergravity. This supergravity multiplet contains five two-form fields with anti-self-dual field strengths in addition to the graviton and fermions. This theory is anomalous unless $N=21$, a fact that was derived long ago \cite{Townsend:1983xt}. There is a simple way to understand this result starting from type IIB supergravity. If one considers compactification on a K3 manifold, one ends up with a theory whose massless sector in six dimensions consists precisely of $(2,0)$ supergravity coupled to 21 tensor multiplets. Moreover, there is a $SO(21,5; \IZ)$ duality group that acts linearly on the $21+5 =26$ string charges. This is analogous to the $SL(2,\IZ)$ duality group in ten dimensions. This is the story at generic points in the moduli space, which has $21\times 5 = 105$ dimensions.  There are singularities in the moduli space at fixed points of the duality group. Many of these correspond to degenerations of two cycles of the K3, and these singularities have an ADE classification. At these singularities a collection of 2-cycles, whose intersections are encoded in an ADE Dynkin diagram, collapse to a point. The strings associated with positive roots, which we have postulated, arise from D3-branes that wrap appropriate two cycles. As the cycles collapse their tensions approach zero.

Suppose we describe the scaling of these two-cycles to zero size in terms of a parameter $\l \to 0$. Then, suppose we redefine all distance scales in the problem by $\l^{-1}$, so that the size of the
cycles remains finite in the limit. This has the effect of decompactifying the K3, decoupling gravity, and sending any other two cycles to infinity. In the limit we are left with a decoupled nongravitational theory in six dimensions, which is precisely the theory that we want. This type of analysis goes by the name of ``geometric engineering'' and has been studied extensively, most recently in the context of F-theory. The name ``local model'' is sometimes used in that context to describe the decoupled nongravitational theory. If one only wants to formulate decoupled $(2,0)$ theories, and not to derive them by this decoupling procedure, there is no compelling reason to exclude ADE algebras with ranks that exceed 21. Indeed, AdS/CFT requires that any rank should be possible. These theories have global anomalies that can be measured by coupling them to background $USp(4)$ gauge fields and background gravitational fields. Since they are anomalies of global symmetries, they do not render the theories inconsistent. Rather, they provide useful topological probes of the physics.


The message is that the six-dimensional superstrings that were postulated in the previous subsection can be regarded as originating from D3-branes wrapping 2-cycles. In the K3 context these strings belong to an $SO(21,5; \IZ)$ multiplet of six-dimensional superstrings, which carry 26 charges. Some of the other strings in this multiplet derive from the $(p,q)$ strings in ten dimensions localized at points in the K3. Being related by the duality group, all of these strings should have much the same structure. In the decoupling limit $\l \to 0$, only the ones associated to positive roots (and their orientation reversals, which correspond to negative roots) of an ADE Lie algebra should survive.

\subsection{String junctions and string webs}

As pointed out in \cite{Lee:2006gqa}, when the rank of the $(2,0)$ theory is greater than one, the strings can form junctions and webs much as they can in ten dimensions. The first example
is the $A_2$ theory, which has three strings with charges (1,0), (0,1) and (1,1). In this case there are two possible string junctions, which are related by orientation reversal. Each of them involves one string of each type. String junctions span a plane (if the three M5-branes are not collinear), and thus the configuration is quarter BPS. Even this simple case can lead to complex string networks. For example, one could tile a plane with a network of hexagons. There are also networks that are not planar and preserve less supersymmetry.

In the $A_{N-1}$ case there are $N(N-1)/2$ strings that arise from M2-branes connecting pairs of
M5-branes. The number of distinct string junctions is $2\times N(N-1)(N-2)/6$. The strings at a
junction consist of the three strings associated with any triple of M5-branes. The factor of two
accounts for orientation reversal and the second factor correspond to choosing any three of the
$N$ M5-branes. One of the challenges in this subject is to account for the number of degrees of
freedom in the SCFT that is predicted by the dual AdS construction. In the simpler example of
$SU(N)$ MSYM the answer is $N^2 -1$, which is well understood from both the CFT and
AdS viewpoints. In the case of the $A_{N-1}$ $(2,0)$ theories, the AdS prediction at leading order
in $N$ is $N^3/3$ \cite{Klebanov:1996un}. It has been noted in \cite{Bolognesi:2011rq} that this corresponds to the
number of three-string junctions. In fact, they even point out that if one adds $N(N-1)$
as the contribution of individual strings to the number of three-string junctions, then the subleading powers of $N$ also agree. The anomaly-based derivation \cite{Harvey:1998bx} is more compelling, but this one (if correct) has identified the degrees of freedom that are being counted.

\subsection{$SU(2)$ MSYM theory in five dimensions}

Let us consider the simplest nontrivial $(2,0)$ theory, the $A_1$ theory on the Coulomb branch, compactified on a circle. This is the theory associated
to two parallel M5-branes. It consists of a single string, whose tension $T$ is equal to the separation of the two M5-branes times the M2-brane tension. Compactification of one of the dimensions on a circle of radius $R$ is expected to give five-dimensional $SU(2)$ MSYM with coupling constant $g_{YM}^2 \sim R$ in the Coulomb phase.
Since all of the configurations that we will discuss are BPS, it is automatic that the masses
and tensions agree with what is computed in the gauge theory. The nontrivial requirement is that the complete list is reproduced. There is also the question whether the five-dimensional MSYM
theory exists as a perturbatively finite quantum theory, as was conjectured by Douglas \cite{Douglas:2010iu}. In a {\em tour de force}, Douglas and collaborators found that five-dimensional MSYM is ultraviolet divergent at six loops \cite{Bern:2012di}. In principle, the $(2,0)$ construction provides its UV completion.

\subsubsection{Half-BPS point particles}

There are several type of multiplets that appear: Massless states arise as the dimensional reduction of the massless tensor multiplet in six dimensions. This gives a massless vector multiplet in five dimensions. Massive states can arise either as string winding modes or as Kaluza--Klein
excitations. These are both half-BPS. Just as we found in nine dimensions, there are two possible half-BPS massive supermultiplets in five dimensions. The massive little group in five dimensions is
the same as the massless one in six dimensions, namely $SU(2) \times SU(2)$. The Kaluza--Klein excitation supermultiplets are identical to the six-dimensional tensor multiplet, namely $ (3,1) + 4(2,1) + 5(1,1).$ This multiplet is chiral. The half-BPS supermultiplets associated with winding, on the other hand, are non-chiral. Just as in ten dimensions, the relative chirality of the left and right movers is reversed for winding states so that the multiplet has the content
\begin{equation} [(2,1) +2(1,1)] \times [(1,2) +2(1,1)] =  (2,2) + 2(2,1) + 2(1,2) + 4(1,1). \end{equation}
This is the same as a massless vector supermultiplet in six dimensions.


The winding number $W$ is a gauge charge, since it couples to the massless vector field. The choices $W=\pm 1$ give the charged members of the broken $SU(2)$ gauge multiplet. The mass of a winding state is $2\pi R |W|$. Winding states with $|W| >1$ would give multiply-charged particles, which we certainly do not want to exist as stable particles. They are at threshold for decay into $W$ singly-charged particle. Therefore it must be the case that the spectrum does not contain such multiply-charged states.

The conserved momentum $K$ is associated to a global $U(1)$ symmetry. This is to be contrasted with gravity theories where the metric provides a gauge field that couples to the KK charge. The masses of these states are $K/R$. The ones with $|K| >1$ are also at threshold for decay into singly charged states, just like the winding states. Being KK modes, the requisite bound states should exist for all $K$. The $K=\pm 1$ states correspond to instanton particles in the five-dimensional gauge theory \cite{Kim:2011mv}. More precisely, they correspond to normalizable wave functions on the single instanton moduli space. They can be constructed in the gauge theory by the standard instanton construction in the four spatial dimensions. Multi-instanton configurations exist, too, and account for the higher KK modes.

\subsubsection{Quarter-BPS point particles}

There should also be quarter-BPS states that carry both electric charge and instanton charge. These states are analogous to the quarter-BPS states in nine dimensions that were discussed in Sect. 2.4.
As in that case, these would arise as excitations of the self-dual string wrapped on the circle that are purely left-moving or right-moving. Recall that the relevant formula is $|WK| =N$, where the left-movers, say, are in the ground state and the right-movers are in the $N$th excited level.

A specific proposal for the CFT that describes the self-dual string was made by Dijkgraaf, Verlinde, and Verlinde in \cite{Dijkgraaf:1996cv} and a somewhat different one was made by me in \cite{Schwarz:1996pi}. One way to settle the question is to carry out an index computation that characterizes the spectrum of string excitations. Such computations were performed in \cite{Kim:2011mv}. The result found there, specifically eq.~(4.6), appears to favor the DVV proposal. However, in my opinion, the question deserves further study.

The DVV proposal corresponds to the quantization of a six-dimensional GS superstring. This is a consistent classical theory, but -- in contrast to ten-dimensional superstrings -- it suffers from a conformal anomaly. In light-cone gauge this is manifested as a breakdown of Lorentz invariance. The proposal in \cite{Schwarz:1996pi} is that the CFT is the same as that of a ten-dimensional superstring at a $\IC^2/\IZ_2$ orbifold singularity, which is an $A_1$ singularity, in the twisted sector. The considerations in Sect. 4.4 seem to support this picture.  Therefore further index calculations are warranted.

\subsubsection{Half-BPS strings}

Finally, the five-dimensional MSYM theory contains a half-BPS string. This is the original unwrapped string itself, now in the five-dimensional theory. Its tension remains $T$, of course. It couples magnetically to the massless vector field. Note that in six dimensions this string carries a self-dual charge. This string can be constructed in the gauge theory by the four-dimensional 't Hooft--Polyakov monopole construction \cite{Boyarsky:2002ck}. It extends along the direction that is not utilized in the construction. The string can also be wound helically along the circle, as we have discussed. In this case it carries an electric charge density $\r$ per unit length. In this case the string is still half-BPS and its five-dimensional energy density is $\sqrt{1+ (2\pi R \r)^2}T$. Lee and Yee generalized the monopole construction to incorporate this charge density \cite{Lee:2006gqa}.

\subsection{The unbroken gauge symmetry phase}

Another interesting issue concerns the limit in which the broken gauge symmetry is restored. In this limit the M5-branes become coincident and the strings become tensionless. The more relevant fact is that their energy density vanishes, but tension and energy density are the same for these strings. In five dimensions the string is given by a monopole construction, and it is known what happens to monopoles when the scalar vacuum expectation value that is responsible for the symmetry breaking is turned off. In this limit the mass of the monopole goes to zero. However, at the same time it becomes more and more delocalized. This is the reason that one does not include massless monopoles and dyons in the unbroken symmetry phase of ${\cal N} =4$ SYM, for example. By the same token, one should not expect there to be a local formulation of six-dimensional $(2,0)$ theories in the unbroken symmetry phase in terms of tensionless self-dual strings. Thus, the considerations of this paper, which focus on the properties of strings, may not be directly relevant to understanding these theories in the unbroken symmetry phase. We know from the work of Argyres and Douglas that there are strongly coupled four-dimensional ${\cal N} =2$ conformal field theories in which mutually nonlocal point particles can be simultaneously massless \cite{Argyres:1995jj}. Perhaps the six-dimensional conformal $(2,0)$ theories are somewhat analogous field theories inasmuch as they are strongly coupled, and the tensionless strings are nonlocal because they are self-dual.

The construction of sensible quantum theories of tensionless strings is very tricky, maybe even impossible. While the quantum theory is mysterious, there are a few things that one can say about classical tensionless strings: in a Lorentz invariant theory such a string should travel with the speed of light, the infinite Lorentz contraction forces its position to be restricted to a plane orthogonal to the direction of its motion, and the infinite time dilatation forces its configuration in this plane to be completely rigid.


\section{Conclusion}

The duality between type IIB superstring theory compactified on a circle and M-theory compactified on a torus provides a useful tool for relating the gauge theories associated with multiple parallel D3-branes (MSYM theories in four dimensions) and those associated with multiple parallel M5-branes ($(2,0)$ theories in six dimensions).
Even though the $(2,0)$ SCFTs are strongly interacting and rather elusive, one can identify the BPS branes of these theories on the Coulomb branch. In addition to massless tensor multiplets associated with the Cartan subalgebra of an ADE Lie algebra, they consist of half-BPS self-dual strings associated with the nonzero roots of the Lie algebra. In addition, these strings can form junctions and networks, which preserve less supersymmetry, depending on how they are configured. This understanding is helpful for understanding the BPS branes that are present when a $(2,0)$ theory is compactified. The quarter-BPS spectrum of five-dimensional MSYM theory deserves further study in order to determine the world-sheet conformal field theory of the self-dual string.

\section*{Acknowledgment}

This work was supported in part by the U.S. Dept. of Energy under
Grant No. DE-FG03-92-ER40701.  JHS acknowledges the hospitality of the Aspen Center for Physics during the completion of this paper. The ACP is supported by the National Science Foundation Grant No. PHY-1066293. He also acknowledges discussions with J. Distler, K. Lee, J. Maldacena, C. Vafa, and A. Vainshteyn.

\end{document}